\documentclass[3p,times,procedia]{elsarticle}
\usepackage{nupha_ecrc}

\volume{00}

\firstpage{1}

\journalname{Nuclear Physics A}

\runauth{}

\jid{nupha}

\jnltitlelogo{Nuclear Physics A}

\usepackage{amssymb}

\usepackage[figuresright]{rotating}

\begin{document}

\begin{frontmatter}


\title{Towards a reliable lower bound on the location of the critical endpoint}
\author{M. Giordano}
\author{K. Kapas}
\author{S.D. Katz}
\author{D. Nogradi}
\author[speaker]{A. Pasztor}


\dochead{}


\fntext[speaker]{speaker: apasztor@bodri.elte.hu}
\address{ELTE E\"otv\"os Lor\'and University, Institute for
  Theoretical Physics, P\'azm\'any P.\ s.\ 1/A, H-1117, Budapest,
  Hungary}

\begin{abstract}
We perform the first direct determination of the position of the leading singularity
of the pressure in the complex chemical potential $\mu_B$ plane in lattice QCD using 
numerical simulations with 2-stout improved rooted staggered fermions. This provides 
a direct determination of the radius of convergence of the Taylor expansion of the
pressure that does not rely on a finite-order truncation of the expansion.
The analyticity issues in the complex $\mu_B$ plane of the grand canonical partition function
of QCD with rooted staggered fermions are solved with a careful redefinition of the fermion
determinant that makes it a polynomial in the fugacity on any finite lattice, without changing
the continuum limit of the observables. 
By performing a finite volume scaling study at a single coarse lattice spacing, we 
show that the limiting singularity is not on the real line in the thermodynamic limit, 
thus showing that the radius of convergence of the Taylor expansion gives a lower 
bound on the location of a possible phase transition. 
In the vicinity of the crossover temperature at zero chemical potential, 
the radius of convergence turns out to be $\mu_B/T \approx 2$ and roughly temperature independent.
\end{abstract}


\end{frontmatter}


\section{Introduction}
The theoretical study of the phase diagram of QCD in the 
temperature ($T$)-baryon chemical potential ($\mu_B=3\mu_q=3\mu$) plane is of considerable interest
for the physics of high energy nuclear collisions and for the understanding of the early
stages of the universe.
It is by now established through first-principles lattice QCD calculations that at
$\mu_B=0$ there is an analytic crossover~\cite{Aoki:2006we} at a temperature~\cite{Borsanyi:2010bp} of $T_c\approx 150-160~{\rm MeV}$.
It is further conjectured~\cite{Fukushima:2013rx} that in the 
$(T,\mu_B)$ plane there is a line of crossovers, departing from
$(T_c,0)$, that eventually turns into a line of first-order phase
transitions. The point $(T_{\rm CEP},\mu_{\rm CEP})$ separating
the two lines is known as the critical
endpoint (CEP), and the transition is expected to be of second
order there. Full confirmation of this picture is hindered by 
the notorious sign problem, which prevents direct simulation at $\mu_B>0$,
but several extrapolation techniques have been developed to study QCD at finite small densities. These include Taylor
expansion at $\mu_B=0$~\cite{Allton:2002zi,Bazavov:2017dus}, analytic continuation from imaginary
chemical potential~\cite{deForcrand:2002hgr,Bonati:2018nut,Borsanyi:2018grb}, and reweighting~\cite{Hasenfratz:1991ax,Fodor:2001pe}.  
The basic idea of these methods is to reconstruct the behavior of the
theory at real $\mu_B>0$ by extrapolating from zero or purely imaginary $\mu_B$, where
the sign problem is absent. 
In the context of these extrapolation methods, one of the most sought-after quantities in the
finite temperature lattice QCD community is the radius of convergence of the Taylor expansion of the pressure around $\mu_B=0$~\cite{Bazavov:2017dus,DElia:2016jqh,Fodor:2019ogq}. 
The reason for this is twofold. First, the radius of convergence gives information about the location of the supposed CEP.
More precisely, the radius of convergence gives a rigorous lower bound on the
location of possible phase transitions. Moreover, if the leading singularity in the thermodynamic 
limit is on the real axis then it corresponds to a genuine phase transition.
Second, one of the most important inputs of lattice QCD to the phenomenology of heavy ion 
collisions is the equation of state, including its extrapolation to finite $\mu_B$. 
Knowledge of the radius of convergence would tell us how far such an extrapolation is trustworthy.
Even if the lower bound on the CEP location happens to be not very stringent, this is still
very useful information, since viscous hydrodynamics simulations of heavy ion collisions
at intermediate energies already explore a very wide range of temperatures and baryo-chemical 
potentials. Currently available lattice estimates of the radius of 
convergence are obtained using
the first few Taylor coefficients, and with their statistical significance
strongly deteriorating with order, it would seem 
rather unlikely that the convergence of the estimators of the radius of convergence
can be demonstrated in the near future. This contribution introduces a novel method,
with which this problem may be circumvented. More details can be found in Refs.~\cite{Giordano:2019slo} and \cite{LongPaper}.

\section{Lee-Yang zeros and the exact asymptotics of the Taylor expansion in a finite volume}
In a “typical” statistical  mechanical  system the grand canonical partition function is, up to a non-vanishing factor, a polynomial of
the fugacity $e^{\mu/T}$ (the Lee-Yang polynomial). In such a situation, the zeros of this polynomial, the so-called Lee-Yang zeros determine the analytic properties
of the pressure $\propto \log Z$. Namely at each Lee-Yang zero, the pressure develops a logarithmic branch point singularity. 
The Lee-Yang zeros have some simple symmetry properties, namely, they have a $\mu \to \mu^*$ symmetry due to
the fact that the coefficients of the Lee-Yang polynomial are the canonical partition functions, i.e. they are real numbers $Z_n>0$. 
Second, in QCD-like theories, due to CP symmetry they also have a $\mu \to -\mu$ symmetry. This means in particular the in the Taylor expansion 
$\log Z = \sum_n c_n \mu^{2n}$ the high order terms are dominated by 4 logarithmic singularities (Lee-Yang zeros) at the same distance from $\mu=0$.
This is enough information to calculate the exact asymptotic behavior of the Taylor expansion in any finite volume as~\cite{Giordano:2019slo}: $c_{k} \sim \frac{-2}{k} \frac{\cos(2k\theta)}{r^{2k}} \quad {\rm{as}} \quad k \to \infty$,
where $r$ and $\theta$ are the polar coordinates of an arbitrary member of the closest lying quartet of Lee-Yang zeros. This formula has some interesting immediate consequences, namely:
i) The ratio estimator, widely used in the literature to estimate the radius of convergence can never converge in any finite volume, where the lattice simulations
are actually done; ii) Even if a given gauge ensemble is large enough to determine $r$ and $\theta$ accurately, the errorbars of the high order coefficients will 
always blow up (as can be checked by applying linear error propagation applied to the above formula) iii) The fluctuations of asymptotically high order are strongly 
correlated on any given gauge ensemble, since they are all given by $r$ and $\theta$. The last point in turn implies that if the correlations between the $c_n$ 
are kept, even if they have $>100\%$ errorbars, the leading Lee-Yang zero position may still be recovered. This was explicitly demonstrated on a small toy lattice in
~\cite{Giordano:2019slo}. How the leading Lee-Yang zero can be recovered from the Taylor coefficients is also explained in~\cite{Giordano:2019slo}, where improved
estimators of the radius of convergence are constructed, that take into account the exact asymptotic form of the Taylor coefficients discussed above.

The main take-away of this discussion is that what allows one to reconstruct the leading Lee-Yang zero in the first place is the strong correlations between the
large order Taylor coefficients, which make possible a very striking cancellation of statistical errors in certain combinations of them, corresponding to the
position of the Lee-Yang zeros. This means that in order to eventually calculate the radius of convergence in continuum QCD, one is probably better served by
calculating the radius of convergence on a fixed lattice spacing and volume first, and only later tackle the continuum and infinite volume limits at the level
of the radius of convergence itself. Unfortunately, as we will soon discuss, this is not possible with the numerically cheapest formulation of lattice QCD,
rooted staggered fermions, as the partition function with this discretization has some spurious singularities, discussed below.

\section{Spurious non-analyticities with rooted staggered fermions and how to remove them}
The staggered discretization corresponds to 4 degenerate flavors of fermions in the continuum limit. In order to describe
$N_f=1$ or $2$, the so-called rooting trick is used, where one writes the partition function as:
$Z = \int \mathcal{D} U (\det M)^{N_f/4} e^{-S_G[U]}$, 
where $\det M$ is the determinant of the staggered Dirac matrix. When the baryon chemical potential $\mu_B>0$ the determinant is
a complex number. This is the famous sign problem of lattice QCD. In order to define the $N_f=2$ determinant, one then has 
to choose a branch of the square root function.  The standard methods of analytic continuation and Taylor expansion use here a branch of the
square root that continuously connects to the positive square root at $\mu_B=0$. While this choice is expected to be perfectly
fine near the continuum limit, at a finite lattice spacing it has the non desirable property that the partition function develops 
a square root type branch point whenever the determinant has a zero on a given gauge configuration. A natural consequence of this is
that the partition function is non-analytic everywhere on the support of the probability density of the zeros of the staggered determinant.
In such a case the radius of convergence of the pressure is bounded from above by the point of this support closest to $\mu=0$.
This is a cut-off effect that at any finite lattice spacing potentially reduces the radius of convergence to a smaller, unphysical value, as the radius of convergence is no longer given by a partition function zero, but rather the closest zero of the staggered determinant on
the entire ensemble. With the standard methods, the obvious solution to this problem is to calculate the continuum  limit  of the 
individual $c_n$ first, and only later study the high order behavior of the series.
Unfortunately this destroys the strong correlations
present between the $c_n$ on a single ensemble, which allows one to reconstruct the position of the leading Lee-Yang zero in the
first place.

\begin{figure*}[t]
  \centering
  \hspace{-0.35cm} 
  \rotatebox{-90}{\includegraphics[width=0.32\textwidth]{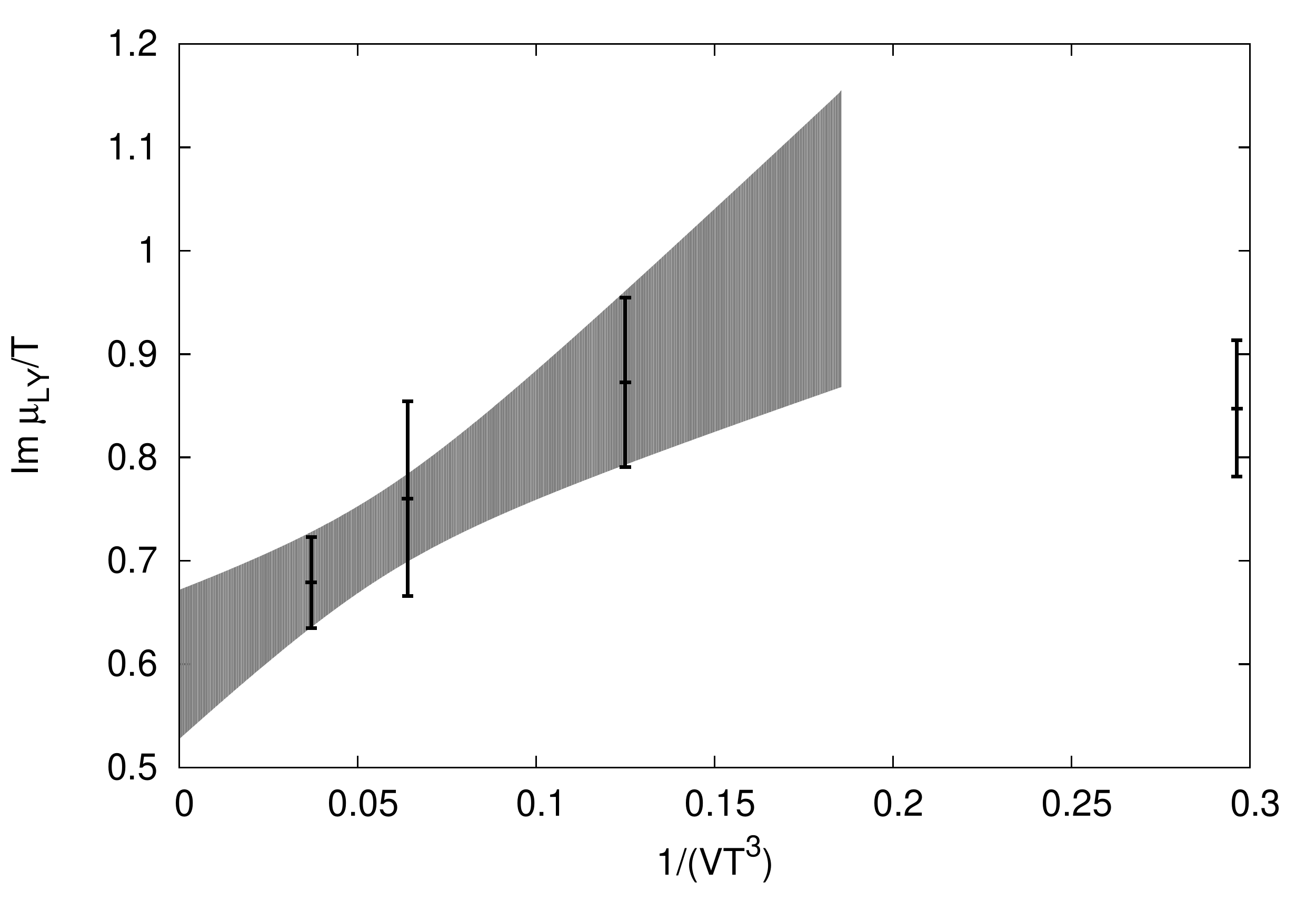}}   \hspace{-0.35cm}
  \rotatebox{-90}{\includegraphics[width=0.32\textwidth]{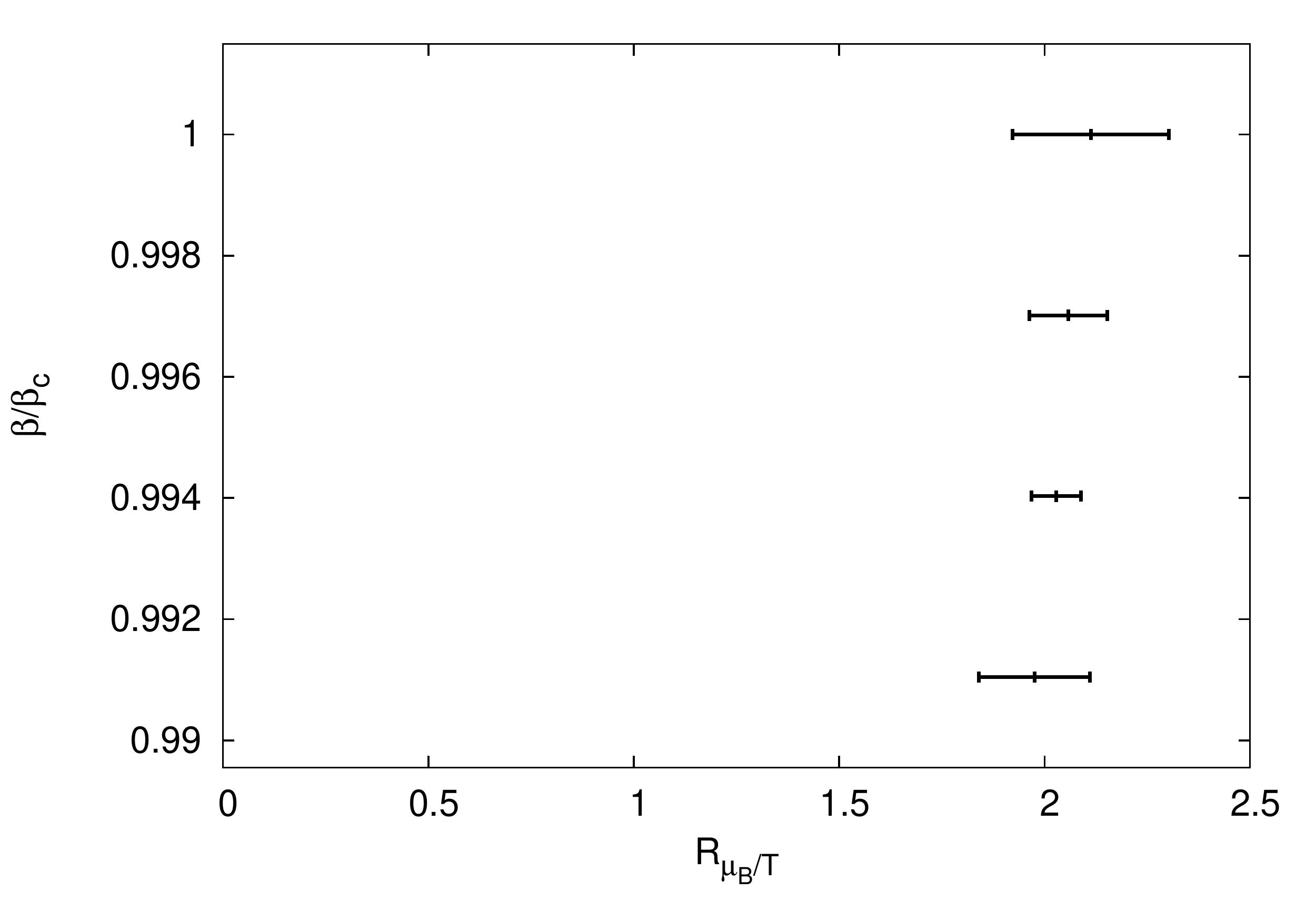}}    
  \hspace{-0.45cm}
    \caption{Left: Imaginary part of the leading Lee-Yang zero against inverse volume at the $\mu=0$ crossover temperature (in units of the inverse gauge coupling) $\beta=\beta_c=3.35$. A linear fit in $1/V$ for the 
                   volumes $8^3, 10^3$ and $12^3$ is also shown. The extrapolation looks similar for all temperatures considered in this work.
           Right: The radius of convergence extrapolated to infinite volume for all the
                  temperatures studied in this work (expressed in terms of the bare gauge coupling).} 
  \label{fig:technicalities}
\end{figure*}

We now proceed to describe a possible resolution to the problem discussed above. 
This involves introducing a new definition of the $N_f=2$ determinant with staggered fermions, that does not
involve square roots.
It has been shown in Ref.~\cite{Hasenfratz:1991ax}
that the staggered determinant on a fixed gauge configuration can be calculated for any complex value of the quark chemical potential $\mu$ as
$    \det M(\mu/T)_{N_f=4} = e^{-3 V \mu/T } \prod_{i=1}^{6 V} \left(
  \xi_i - e^{\mu/T} \right)$,
where $V$ is the spatial volume in lattice units,  and $N_s$ is the spatial linear
size of the lattice in lattice units and the $\xi_i$ are eigenvalues of the so-called reduced matrix.  
The reduced matrix and therefore the eigenvalues $\xi_i$ only depend on the gauge configuration and 
not on the chemical potential. Near the continuum limit, the fact that the staggered operator describes 4 flavors of
fermions manifests itself in a near degeneracy of the eigenvalues of the original Dirac operator $D$ as well as those of the
reduced matrix~\cite{LongPaper}. One can then define the two-flavor determinant by grouping closely lying pairs of eigenvalues
of the reduced matrix and substituting each pair $(\xi_{p_1},\xi_{p_2})$ with a suitably defined ``average eigenvalue'' $\tilde{\xi}_p$
and defining the two-flavor determinant as $\det M(\mu/T)_{N_f=2} = e^{-3 V \mu /(2T) } \prod_{p=1}^{3 V} \left(\tilde{\xi}_p - e^{\mu/T} \right)$
with the sum now running over the pairs. This way the definition of the 2 flavor determinant does not involve any square roots
and therefore no branch point singularities. This then guarantees that the radius of convergence will again be given by the partition function zero closest to $\mu=0$. With this new definition in hand we can embark on the calculation of the leading Lee-Yang zero. To test the idea
numerically we conducted simulations with $N_t=4$ 2stout improved staggered fermions with lattice spatial volumes of $6^3,8^3,10^3$ and $12^3$ and calculated the position of the
leading Lee-Yang zero for all cases. We also performed a simple infinite volume extrapolation with a linear ansatz in $1/V$. The nonzero
infinite volume limit of the imaginary part indicates that the radius of convergence of the pressure is not given by a genuine phase transition,
but rather by a complex singularity. Our numerical results are summarized briefly in Fig. 1. On the left we show an infinite volume extrapolation of the imaginary part of the leading Lee-Yang zero at a specific temperature, demonstrating that the 
leading singularity is complex, and therefore that the radius of convergence is not given by a phase transition. On the right, we show the infinite volume
extrapolation of the radius of convergence as a function of the temperature. For more details, we refer the reader to Ref.~\cite{LongPaper}.
Extending such a study to finer lattices is challenging. However, our approach has the conceptual advantage over the standard definition of staggered rooting
of dealing with an actual Lee-Yang polynomial at a finite lattice spacing, instead of a function that is non-analytic in dense regions of the complex chemical potential plane.
This makes the challenge worth the pursuit.

\section*{Acknowledgements}
This work was partially supported by the Hungarian National
Research, Development and Innovation Office - NKFIH grant KKP126769 
and by OTKA under the grant OTKA-K-113034. A.P. is supported by 
the J\'anos Bolyai Research Scholarship of the Hungarian Academy of 
Sciences and by the \'UNKP-19-4 New National Excellence Program of 
the Ministry for Innovation and Technology.


\begin{thebibliography}{10}
\expandafter\ifx\csname url\endcsname\relax
  \def\url#1{\texttt{#1}}\fi
\expandafter\ifx\csname urlprefix\endcsname\relax\def\urlprefix{URL }\fi
\expandafter\ifx\csname href\endcsname\relax
  \def\href#1#2{#2} \def\path#1{#1}\fi

\bibitem{Aoki:2006we}
Y.~Aoki, G.~Endrodi, Z.~Fodor, S.~D. Katz, K.~K. Szabo, {The Order of the
  quantum chromodynamics transition predicted by the standard model of particle
  physics}, Nature 443 (2006) 675--678.
\newblock \href {http://arxiv.org/abs/hep-lat/0611014}
  {\path{arXiv:hep-lat/0611014}}, \href {http://dx.doi.org/10.1038/nature05120}
  {\path{doi:10.1038/nature05120}}.

\bibitem{Borsanyi:2010bp}
S.~Bors{\'a}nyi, et~al., {Is there still any $T_c$ mystery in lattice QCD?
  Results with physical masses in the continuum limit III}, JHEP 1009 (2010)
  073.
\newblock \href {http://arxiv.org/abs/1005.3508} {\path{arXiv:1005.3508}},
  \href {http://dx.doi.org/10.1007/JHEP09(2010)073}
  {\path{doi:10.1007/JHEP09(2010)073}}.

\bibitem{Fukushima:2013rx}
K.~Fukushima, C.~Sasaki, {The phase diagram of nuclear and quark matter at high
  baryon density}, Prog. Part. Nucl. Phys. 72 (2013) 99--154.
\newblock \href {http://arxiv.org/abs/1301.6377} {\path{arXiv:1301.6377}},
  \href {http://dx.doi.org/10.1016/j.ppnp.2013.05.003}
  {\path{doi:10.1016/j.ppnp.2013.05.003}}.

\bibitem{Allton:2002zi}
C.~R. Allton, S.~Ejiri, S.~J. Hands, O.~Kaczmarek, F.~Karsch, E.~Laermann,
  C.~Schmidt, L.~Scorzato, {The QCD thermal phase transition in the presence of
  a small chemical potential}, Phys. Rev. D66 (2002) 074507.
\newblock \href {http://arxiv.org/abs/hep-lat/0204010}
  {\path{arXiv:hep-lat/0204010}}, \href
  {http://dx.doi.org/10.1103/PhysRevD.66.074507}
  {\path{doi:10.1103/PhysRevD.66.074507}}.

\bibitem{Bazavov:2017dus}
A.~Bazavov, et~al., {The QCD Equation of State to $\mathcal{O}(\mu_B^6)$ from
  Lattice QCD}, Phys. Rev. D95~(5) (2017) 054504.
\newblock \href {http://arxiv.org/abs/1701.04325} {\path{arXiv:1701.04325}},
  \href {http://dx.doi.org/10.1103/PhysRevD.95.054504}
  {\path{doi:10.1103/PhysRevD.95.054504}}.

\bibitem{deForcrand:2002hgr}
P.~de~Forcrand, O.~Philipsen, {The QCD phase diagram for small densities from
  imaginary chemical potential}, Nucl. Phys. B642 (2002) 290--306.
\newblock \href {http://arxiv.org/abs/hep-lat/0205016}
  {\path{arXiv:hep-lat/0205016}}, \href
  {http://dx.doi.org/10.1016/S0550-3213(02)00626-0}
  {\path{doi:10.1016/S0550-3213(02)00626-0}}.

\bibitem{Bonati:2018nut}
C.~Bonati, M.~D'Elia, F.~Negro, F.~Sanfilippo, K.~Zambello, {Curvature of the
  pseudocritical line in QCD: Taylor expansion matches analytic continuation},
  Phys. Rev. D98~(5) (2018) 054510.
\newblock \href {http://arxiv.org/abs/1805.02960} {\path{arXiv:1805.02960}},
  \href {http://dx.doi.org/10.1103/PhysRevD.98.054510}
  {\path{doi:10.1103/PhysRevD.98.054510}}.

\bibitem{Borsanyi:2018grb}
S.~Bors{\'a}nyi, Z.~Fodor, J.~N. G{\"u}nther, S.~K. Katz, K.~K. Szab{\'o},
  A.~P{\'a}sztor, I.~Portillo, C.~Ratti, {Higher order fluctuations and
  correlations of conserved charges from lattice QCD}, JHEP 10 (2018) 205.
\newblock \href {http://arxiv.org/abs/1805.04445} {\path{arXiv:1805.04445}},
  \href {http://dx.doi.org/10.1007/JHEP10(2018)205}
  {\path{doi:10.1007/JHEP10(2018)205}}.

\bibitem{Hasenfratz:1991ax}
A.~Hasenfratz, D.~Toussaint, {Canonical ensembles and nonzero density quantum
  chromodynamics}, Nucl. Phys. B371 (1992) 539--549.
\newblock \href {http://dx.doi.org/10.1016/0550-3213(92)90247-9}
  {\path{doi:10.1016/0550-3213(92)90247-9}}.

\bibitem{Fodor:2001pe}
Z.~Fodor, S.~D. Katz, {Lattice determination of the critical point of QCD at
  finite T and mu}, JHEP 03 (2002) 014.
\newblock \href {http://arxiv.org/abs/hep-lat/0106002}
  {\path{arXiv:hep-lat/0106002}}, \href
  {http://dx.doi.org/10.1088/1126-6708/2002/03/014}
  {\path{doi:10.1088/1126-6708/2002/03/014}}.

\bibitem{DElia:2016jqh}
M.~D'Elia, G.~Gagliardi, F.~Sanfilippo, {Higher order quark number fluctuations
  via imaginary chemical potentials in $N_f=2+1$ QCD}, Phys. Rev. D95~(9)
  (2017) 094503.
\newblock \href {http://arxiv.org/abs/1611.08285} {\path{arXiv:1611.08285}},
  \href {http://dx.doi.org/10.1103/PhysRevD.95.094503}
  {\path{doi:10.1103/PhysRevD.95.094503}}.

\bibitem{Fodor:2019ogq}
Z.~Fodor, M.~Giordano, J.~N. Günther, K.~Kapás, S.~D. Katz, A.~Pásztor,
  I.~Portillo, C.~Ratti, D.~Sexty, K.~K. Szabó, {Trying to constrain the
  location of the QCD critical endpoint with lattice simulations}, Nucl. Phys.
  A982 (2019) 843--846.
\newblock \href {http://dx.doi.org/10.1016/j.nuclphysa.2018.12.015}
  {\path{doi:10.1016/j.nuclphysa.2018.12.015}}.

\bibitem{Giordano:2019slo}
M.~Giordano, A.~P\'asztor, {Reliable estimation of the radius of convergence in
  finite density QCD}, Phys. Rev. D99~(11) (2019) 114510.
\newblock \href {http://arxiv.org/abs/1904.01974} {\path{arXiv:1904.01974}},
  \href {http://dx.doi.org/10.1103/PhysRevD.99.114510}
  {\path{doi:10.1103/PhysRevD.99.114510}}.

\bibitem{LongPaper}
M.~Giordano, K.~Kap\'as, S.~D. Katz, A.~P\'asztor, D.~N\'ogr\'adi, {Radius of
  convergence in lattice QCD at finite $\mu_B$ with rooted staggered fermions,
  }\href {http://arxiv.org/abs/hep-lat/1911.00043}
  {\path{arXiv:hep-lat/1911.00043}}.

\end{thebibliography}

\end{document}